%% file: 0_main.tex
\documentclass[journal]{vgtc}                          

\graphicspath{{figures/}{pictures/}{images/}{./}} 

\usepackage{times}                     

\usepackage{tabu}                      
\usepackage{booktabs}                  
\usepackage{lipsum}                    
\usepackage{mwe}                       
\usepackage{enumitem}

\usepackage{bm}  

\usepackage{tabularx}
\usepackage{array}
\usepackage{makecell}
\usepackage{threeparttable}
\usepackage{multirow}
\usepackage{adjustbox}

\newcolumntype{Y}{>{\centering\arraybackslash}X}

\usepackage{mathptmx}                  

\usepackage{xcolor}

\newcommand{\HDY}[1] 
{\textcolor{black}{#1}}
\newcommand{\ic}[1] 
{\textcolor{black}{#1}}

\newcommand{\factor}[1] 
{\textit{\textbf{#1}}}

\newcommand{\sy}[1] 
{\textcolor{black}{#1}}

\onlineid{0}

\vgtccategory{Research}

\title{Smelling the Way: Olfactory Modulation of Spatial Estimation and Path Integration in Virtual Reality}

\abstract{
Spatial cognition enables individuals to perceive and interpret spatial relationships, estimate locations, and navigate within their surroundings. 
While vision plays a dominant role, other sensory modalities, particularly olfaction, can support spatial processing when visual cues are limited. 
This paper explores the \sy{effects of scent-delivery} within virtual reality (VR) environments \sy{by using fan-mediated olfactory devices.}
We conducted two formal studies to evaluate the \sy{effects on spatial information acquisition.}
Study 1 (N=24) examined the participants' ability to localize a scent source while walking a linear path using two different olfactory devices. 
Study 2 (N=19) employed a triangle completion task, in which participants attempted to return to their starting point under varying rotation angles and olfactory conditions. 
Results indicate that \sy{scent-delivery feedback} influenced spatial behavior in distinct ways across the two studies.
 In Study 1, visual–olfactory misalignment produced systematic directional biases without improving absolute localization accuracy. 
 In Study 2, distance-modulated \sy{scent-delivery feedback} reduced terminal homing error but also increased traveled distance and completion time, suggesting a more deliberate goal-confirmation strategy.
}

\author{%
\authororcid{Siyeon Bak}{0009-0008-8845-2269}, So-Hui Kim,
\authororcid{Junho Kim}{0009-0009-3799-9691}, 
\authororcid{Dongyun Han}{0000-0002-9517-5326},
\authororcid{Sun-Jeong Kim}{0000-0002-8663-4578},  and 
  \authororcid{Isaac Cho}{0000-0003-1582-8428}
}

\authorfooter{
\item
Siyeon Bak, So-Hui Kim, Junho Kim, and Sun-Jeong Kim are with Hallym University
E-mail: 123siyeon84@gmail.com, rlathgml1567@naver.com, rlawnsghdudw@naver.com, sunkim@hallym.ac.kr

\item
Dongyun Han is with Clemson University
E-mail: dongyuh@clemson.edu

\item
Isaac Cho is with Hallym University (corresponding author).
E-mail: drisaaccho@gmail.com
}

\begin{document}



\maketitle


\input{Section/01_intro_new}
\input{Section/02_related_work}
\input{Section/03_experiment_overview}

\input{Section/03_study1}
\input{Section/04_study2}
\input{Section/05_overall_discussion}
\input{Section/06_conclusion}

\section*{Acknowledgments}
This research was supported by Culture, Sports and Tourism R\&D Program through the Korea Creative Content Agency grant funded by the Ministry of Culture, Sports and Tourism in 2026 (Project Number: RS-2026-25525669, Contribution Rate: 33\%), the Institute of Information \& Communications Technology Planning \& Evaluation (IITP) grant funded by the Korea government (MSIT) (RS-2026-25516382, 33\%), and the National Research Foundation of Korea (NRF) grant funded by the Korea government (MSIT) (RS-2026-25492756).






\bibliographystyle{abbrv-doi}

\bibliography{0_ref}

\input{Section/0_supplemental_material}
\end{document}

%% file: Section/01_intro_new.tex
\section{Introduction}

Navigation in the physical world emerges from the integration of multiple sensory modalities.
Humans continuously combine visual, auditory, proprioceptive, vestibular, and olfactory information to estimate location, maintain orientation, and update spatial representations during movement \cite{tolman1948cognitive, golledge1999wayfinding, wolbers2010determines}. 
In contrast, many current immersive virtual reality (VR) systems remain predominantly visual–auditory, often omitting non-visual cues that contribute to natural navigation \cite{laviola20173d, loomis2003visual}. This sensory reduction raises an important question of how the absence or selective reintroduction of non-visual modalities shapes spatial estimation and navigation behavior in immersive environments.

Earlier research in multisensory perception demonstrates that spatial judgments arise from reliability-weighted integration across sensory cues \cite{ernst2002humans, ernst2004merging, fouad2025touching}. Because vision typically provides high spatial precision, it often dominates during cross-modal conflict \cite{colavita1974human}. In VR, visual–auditory inconsistencies have been shown to distort spatial perception and navigation \cite{riecke2009moving}. However, olfaction differs from other spatial modalities. Although it offers lower spatial resolution and slower temporal dynamics, olfactory information supports contextual association and spatial memory \cite{jacobs2012chemotaxis, jacobs2015olfactory}. In the real world, ambient odors can function as spatial landmarks and guide navigation through gradient detection and memory-based localization \cite{porter2007mechanisms}, and recent work suggests that olfactory landmarks can enhance wayfinding performance \cite{schwarz2024memory}.

While olfactory cues have been explored in VR primarily in relation to presence and immersion \cite{munyan2016olfactory, persky2020olfactory, yildirim2025digital}, multisensory interaction \cite{bak2025beyond, han2025if} and reorientation mechanisms \cite{lee2022auditory, han2025if}, their roles in spatial estimation and navigation behavior remain less understood. 
\sy{Although prior behavioral studies have examined human scent-source localization in real-world environments \cite{porter2007mechanisms, jacobs2015olfactory}, it remains unclear how these findings translate to VR, where the physical placement of scent-delivery devices and visual–olfactory spatial misalignment introduce additional constraints.
These VR-specific constraints raise important questions regarding localization accuracy under current display systems\cite{javerliat2022nebula, dobbelstein2017inscent}, the effects of visual–olfactory spatial misalignment on directional estimation, and whether scent delivery influences navigation behavior during homing, in which users return to a previously visited location.}

To address these questions, we conducted two complementary user studies. Study 1 investigated cross-modal spatial alignment and conflict during straight-line navigation, manipulating visual–olfactory congruency and comparing 
\sy{stationary} and \sy{wearable} olfactory delivery systems. Study 2 employed the Triangle Completion Task (TCT) \cite{loomis1993nonvisual, dorado2019homing} to examine how olfactory cues relate to homing precision, travel distance, and completion time under varying rotation angles.  These experiments examine both directional estimation in controlled localization tasks and navigation behavior during path integration. The results indicate that olfactory cues do not primarily function as precise directional signals. Instead, scent appears to influence how users interpret spatial relationships and confirm proximity to goal locations during navigation. In immersive environments, olfactory feedback therefore plays a complementary role, shaping navigation behavior rather than directly determining spatial direction.

Our results show that spatial misalignment between visual and olfactory cues induces systematic directional biases during immersive navigation. They further show that olfactory cues influence navigation behavior by supporting goal confirmation during homing tasks rather than providing precise directional guidance. The results also characterize how different olfactory delivery modalities affect navigation efficiency, behavioral dynamics, and perceived cue reliability, providing insights for the design of olfactory feedback in immersive navigation systems.





%% file: Section/02_related_work.tex
\section{Related Work}
\subsection{Effects of Olfaction in VR}



While visual and auditory modalities remain central to creating immersive VR experiences, recent research has highlighted the importance of incorporating multisensory cues to further enhance realism and deepen user immersion~\cite{ranasinghe2018season, ebrahimi2016empirical, hagelsteen2017faster}. 
Among these, incorporating olfaction into VR has shown particular promise, allowing users to interact more vividly and holistically with virtual environments, extending perception beyond sight and sound~\cite{tsai2021ieeeVRDoes, nakamoto2020virtual, bak2025beyond, han2025if}. 
\sy{Prior work has shown that olfactory cues can enhance realism, presence, and emotional engagement \cite{dinh1999evaluating}.}
\sy{At the same time, however, 
prior work suggests that olfactory cues should be treated as a context-dependent cue rather than a universally beneficial sensory channel \cite{egan2017subjective}. }

Olfaction is a uniquely powerful sense in daily life, known for its strong ability to trigger memory and emotion~\cite{lee2022auditory, chanes2016redefining}. 
When incorporated into VR, olfactory cues can significantly deepen the sense of immersion by enhancing the user’s feeling of presence beyond the visual domain, making the virtual experience feel more tangible and emotionally engaging~\cite{persky2020olfactory}. 
This multisensory enhancement holds transformative potential across a variety of VR applications, including gaming, therapeutic interventions, and memory training.

Olfaction plays a subtle yet impactful role in human navigation and wayfinding by providing environmental cues that support spatial orientation~\cite{porter2007mechanisms}. 
Prior research suggests that olfactory cues can facilitate the formation of cognitive maps and enhance route learning~\cite{jacobs2012chemotaxis}.  In the real world, for example, ambient scents from foods and industrial materials can function as olfactory landmarks, helping individuals encode and recall spatial locations. 
These benefits of olfaction in navigation and wayfinding hold promise for VR as well. 
However, significant challenges remain due to the technical limitations of current scent delivery systems, which may lead to discrepancies in accuracy and fidelity between olfactory experiences in VR and those in the real world. 
As it remains unclear how accurately users can perceive and localize scent sources in VR using existing olfactory devices, this study aims to investigate users’ ability to detect and locate olfactory stimuli within a virtual environment.

\subsection{Olfactory Devices for VR}

 \sy{Various olfactory display designs have been proposed for VR, including stationary, wearable, handheld, and encountered-type systems~\cite{tewell2024review}. }
To enhance realism and immersion in VR experiences, distinct devices have been introduced to deliver multisensory stimuli to users~\cite{dinh1999evaluating, hoffman1998physically, kaur2025senses}.
\sy{There are three primary approaches: stationary, wearable, and handheld olfactory devices.}
The first involves using stationary devices placed in the physical environment to disperse fragrances into the surrounding air~\cite{davis2007smell, brewster2006olfoto}. 
These devices often struggle with ventilating lingering scents and adapting to dynamic virtual scenes, as scent dispersion takes time and may dilute before reaching the user effectively. 
As a result, users may experience weakened scent intensity or a delay in sensing the olfactory cues, reducing the potential for immersion. 

The second approach uses \sy{wearable} olfactory devices~\cite{dobbelstein2017inscent}.
The key components of such devices are typically mounted beneath VR headsets~\cite{gougeh2022multisensory, kato2019wearable, javerliat2022nebula}, although some studies have explored attaching them to VR controllers~\cite{niedenthal2023graspable}.
These devices employ various methods for scent delivery, including vapor diffusion~\cite{ranasinghe2018season, javerliat2022nebula, myung2023enhancing}, as well as heating solid~\cite{dobbelstein2017inscent} or liquid~\cite{covington2018development} scent media. By positioning the scent emitters close to the user's nose, these mounted systems enable more immediate and intense olfactory stimulation, enhancing the realism and immersion of the VR experience~\cite{yanagida2004projection, yamada2006wearable}.
\sy{However, scent delivery methods that operate near the user's nose, such as fan-based vapor diffusion, may inadvertently introduce non-olfactory cues, including airflow and mechanical noise\cite{dobbelstein2017inscent,kato2019wearable,javerliat2022nebula,myung2023enhancing}.}

\sy{Lastly, handheld olfactory devices enable active, on-demand scent interaction by allowing users to intentionally trigger scent delivery \cite{niedenthal2019handheld,niedenthal2023graspable,hezroni2022stories,brument2025scentaur}. 
This design offers greater temporal control over scent presentation but requires users to carry or manipulate a dedicated device, which may increase interaction burden.
}
\sy{In this work, we focus on stationary and head-mounted wearable scent-delivery configurations because they allow us to evaluate how scent-delivery location influences users' ability to localize virtual scent sources in VR.
Handheld devices were excluded because they require users to actively manipulate the scent-delivery device, making them unsuitable for evaluating localization of virtual scent sources in VR.}


\subsection{Wayfinding in VR}


Wayfinding in VR refers to how users perceive, interpret, and navigate spatial layouts within immersive environments~\cite{laviola20173d}. 
It encompasses processes such as recognizing landmarks, maintaining orientation, estimating distances, and making navigational decisions in a virtual space~\cite{elvins1997visfiles, sharma2017influence}. 
However, unlike in the physical world, users in VR often face significant challenges such as distorted distance perception, spatial disorientation, and difficulty forming accurate cognitive maps~\cite{henry1993spatial, loomis2003visual}. 
While several factors contribute to these limitations, such as limited depth cues and constrained fields of view~\cite{bellgardt2017utilizing, chapuis2024comparing}, the absence of non-visual sensory information also plays a particularly critical role in disrupting natural navigation processes that rely on multisensory integration in the real world.

Prior studies have explored the integration of olfactory cues into virtual environments as a means to enhance spatial cognition and improve wayfinding performance~\cite{schwarz2024memory}. 
For example, Jacobs et al.~\cite{jacobs2015olfactory} demonstrated that olfactory cues could serve as spatial landmarks within a virtual environment, helping users encode specific locations and navigate more efficiently.
Their findings suggest that olfactory cues not only enhance users’ orientation and recall of paths but also support egocentric and allocentric navigation strategies, which are essential for building robust mental maps in VR.
Lee et al.~\cite{lee2024evaluation} evaluated how visual, auditory, and olfactory cues could be used as attractors in redirected walking systems to reorient users when navigating VR. 
They found combined use of auditory and olfactory cues further enhanced redirection efficiency while maintaining a high level of user presence and minimizing the perception of manipulation.
However, despite the growing interest in integrating olfactory cues into VR, our understanding of how users perceive the directionality of scents and accurately locate scent sources within virtual environments remains limited. 
Understanding this research gap is important given the inherent challenges in replicating real-world olfactory experiences in VR. 
These challenges arise from various factors, including the technical limitations of current olfactory display systems.
Recognizing these technological constraints, this work 
aims to fill this research gap by systematically investigating how users perceive scent locations during navigation in VR. 


%% file: Section/03_experiment_overview.tex
\section{Experiment Overview}

To investigate the role of 
\sy{scent-delivery feedback} in spatial estimation and navigation behavior, we conducted two complementary experiments that examined cross-modal alignment and path integration at different levels of spatial complexity.
The central research question (RQ) of this work is:
\textit{How does 
\sy{scent-delivery feedback} interact with visual cues to influence spatial estimation and navigation behavior in immersive environments?}


To address this RQ, we designed two studies targeting different aspects of spatial behavior.
Study 1 examines how 
\sy{scent-delivery feedback} influence participants' ability to localize the origin of a scent in VR.
Study 2 extends the investigation to multidirectional navigation tasks that require spatial updating and homing. 
These two experiments are designed to examine the role of 
\sy{scent-delivery feedback} 
across two levels of spatial behavior:

\begin{itemize}[leftmargin=0.1in, noitemsep]
    \item \textbf{Directional localization under cross-modal alignment}, which captures how users interpret spatial relationships between visual and 
    \sy{scent-delivery feedback} when identifying a scent source.
    \item \textbf{Navigation behavior during path integration}, which evaluates how 
    \sy{scent-delivery feedback} influence homing performance when users must update their spatial representation during movement.
\end{itemize}

By examining both localization and path integration, this experimental design provides a more comprehensive understanding of how 
\sy{scent-delivery feedback} contribute to spatial cognition in immersive environments. Specifically, it allows us to assess whether scent primarily affects perceptual localization, navigation behavior, or both when users move through virtual spaces.

%% file: Section/03_study1.tex


\begin{figure*}[t]
  \centering 
  \includegraphics[width=.8\textwidth]{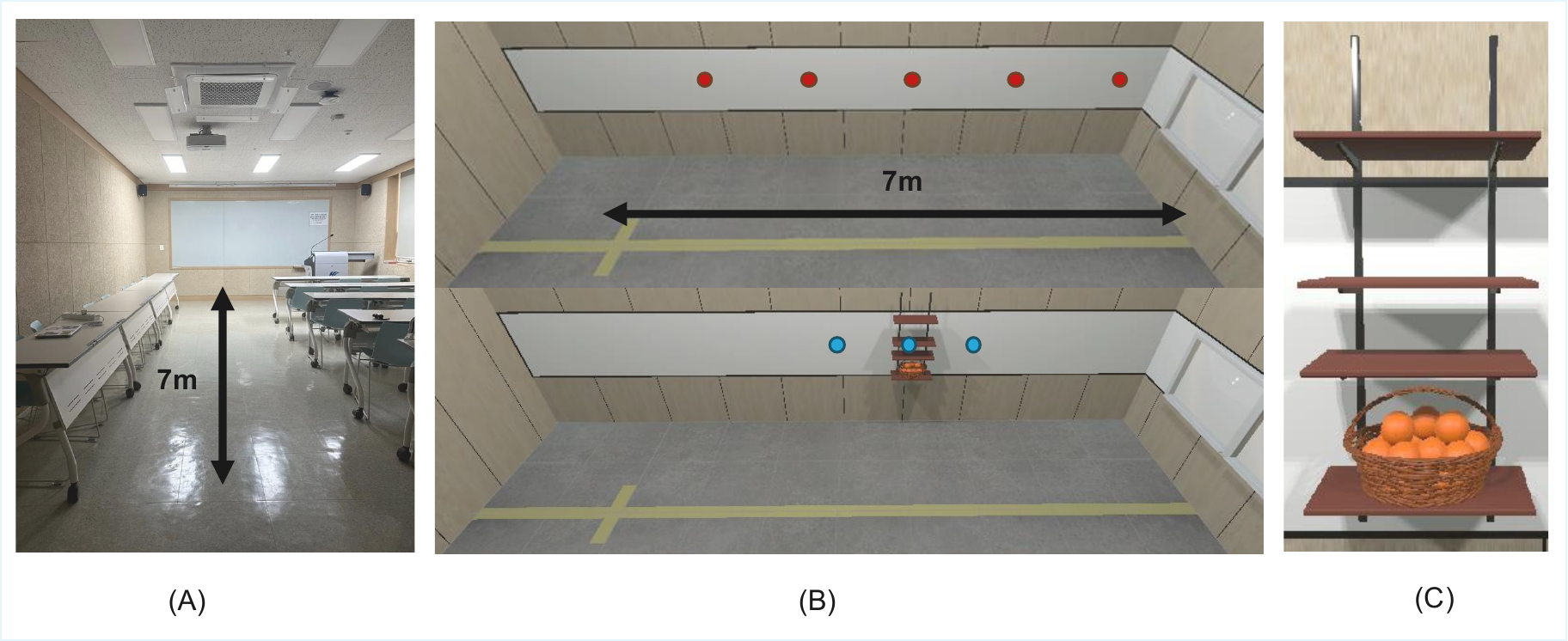}
  \vspace{-0.4cm}
  \caption{(A) shows the physical environment.
  The physical space is mapped one-to-one with the virtual environment, spanning 7 m. (B) shows the corresponding VR scene, which includes a straight line on the floor to guide participants’ walking direction. The virtual orange object (C) is randomly placed at a distance of 3.0 m, 3.5 m, 4.0 m, 4.5 m, or 5.0 m from the starting point (red dots in (B)). The olfactory cue is presented at different positions depending on the condition. For example, in the SLeading and STrailing conditions, the scent is delivered at the first or last blue dot in (B), respectively, when the orange object is located at the center. In the \factor{SOnly} condition, the orange object is hidden.}
  \label{fig_studyspace} 
  \vspace{-0.3cm}
    \end{figure*}

\section{STUDY 1: Cross-Modal Spatial Localization under Visual–Olfactory Alignment}


\subsection{Study Design}
Fig.~\ref{fig_studyspace} illustrates the physical setup and its one-to-one virtual counterpart used in Study 1. Although the classroom contained desks and chairs, we cleared a 5 m × 10 m open area. {The windows remained closed, and the ceiling-mounted air conditioning system remained turned off throughout the study.} Within this space (Fig.~\ref{fig_studyspace}A), a 7 m straight walking path was used for the experiment. 
The VR environment measured the space and reproduced the classroom space at a 1:1 scale, and included yellow straight floor guidelines to guide forward movement and reduce lateral deviation (Fig.~\ref{fig_studyspace}B). Participants stood at the designated starting position and reset the view in Unity to ensure that the virtual and physical spaces were properly aligned.

The study employed a within-subject design with two independent variables: (1) olfactory device type and (2) visual–olfactory spatial alignment condition. A citrus-based orange scent was used as the target stimulus because it was familiar and semantically congruent with the visual orange object used in the task\footnote{The fragrance oil was purchased from \textit{\url{https://www.serimfood.kr/.}}}.  
The virtual orange object was randomly placed at one of the following distances from the starting point: 3.0 m, 3.5 m, 4.0 m, 4.5 m, or 5.0 m. 
\sy{The scent-source position was then defined relative to the virtual orange object according to the experimental condition: either co-located with the object or shifted by ±1 m.} 
When participants reached the scene-source position, the scent was presented once for 3 seconds.

\begin{figure}[t]
  \centering
  \includegraphics[width=0.9\linewidth]{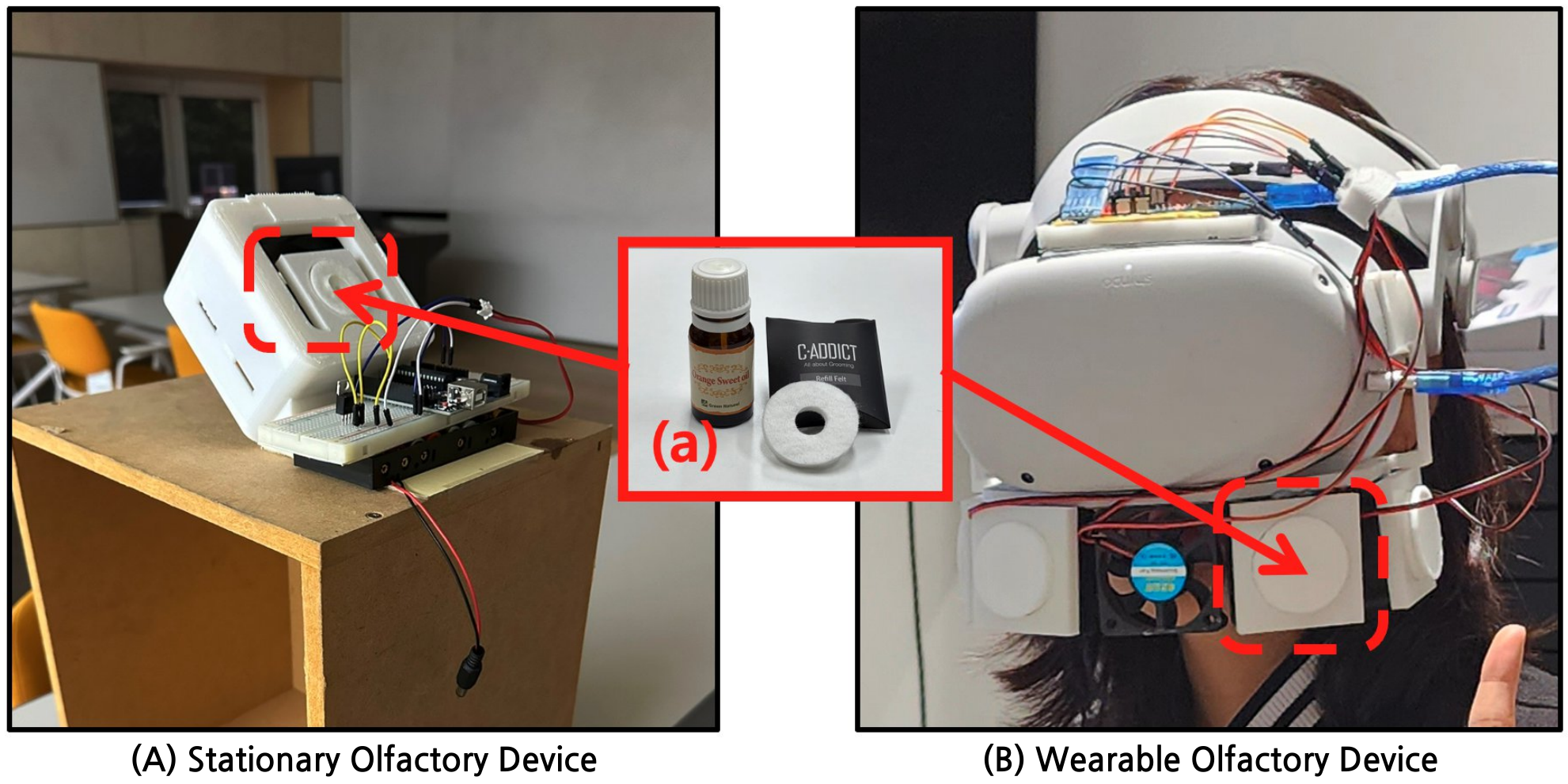}
  \caption{(A) \sy{Stationary} olfactory device and (B) \sy{Wearable} olfactory device used in Study 1.
These devices delivered fragrance using cotton soaked in essential oil (a) and were controlled through a Unity application to either release or stop the scent.
  }\label{fig_olfactoryDevice}
    \vspace{-.5cm}
\end{figure}

To deliver the scent, two olfactory delivery configurations were evaluated: a \sy{wearable} device attached to the HMD and a \sy{stationary} device positioned within the physical environment (Fig.~\ref{fig_olfactoryDevice}).
The \sy{wearable device} configuration provided proximal scent delivery near the participants' nose, whereas the \sy{stationary device} delivered scent from spatially defined locations corresponding to the virtual source. Detailed hardware specifications and operational procedures are described in Section ~\ref{sec_apparatus}. \ic{
In the stationary condition, the physical device was repositioned only to realize the predefined scent-source location under the one-to-one virtual–physical mapping. Rather than shifting the virtual orange object, we repositioned the stationary device to preserve the visual environment across all conditions while manipulating only the spatial relationship between the visual object and the olfactory cue.}

To reinforce the association between scent and visual information, virtual orange objects were placed in the VR environment (Fig.~\ref{fig_studyspace}-C). 
The object’s role was purely experimental control: it provided a salient visual reference that was either redundant with the scent (co-located) or a distractor when spatially misaligned. The object did not provide additional task feedback beyond its static position, and participants were instructed that their task was to locate the scent source position. Using an orange object with an orange odor ensured semantic congruence, so that only spatial congruency was manipulated.

To investigate the effects of 
\sy{scent-delivery feedback} on spatial localization, four experimental conditions were considered by manipulating the alignment between the scent delivery position and the location of the visual object, as follows:

\begin{itemize}[leftmargin=0.15in, noitemsep]
    \item {\textbf{Scent Leading (\textit{SLeading})}}: The scent source is located 1 m ahead of the position where the virtual orange object is located (i.e., at one of 3.0 m, 3.5 m, 4.0 m, 4.5 m, or 5.0 m).    
    
    \item {\textbf{Scent Aligned (\textit{SAligned)}}}: The location where the orange scent is delivered and the virtual object is located are the same.
    
    \item {\textbf{Scent Trailing (\textit{STrailing})}}: The orange scent is delivered at a position 1 m behind the virtual object.
    
     \item {\textbf{Scent Only (\textit{SOnly})}}: This condition provides only the scent, without any visual representation of orange objects in VR. 
\end{itemize}

These four conditions were designed to compare localization accuracy and directional bias across congruent, conflicting, and 
\sy{scent-delivery}-only settings.

\subsection{Study Task and Procedure}

The study took approximately an hour. Upon arrival at the study place, participants were asked to read and sign an informed consent form following the IRB protocol (\sy{HIRB-2024-039-R-R-RE}).
Participants also completed a pre-experiment survey that collected demographic information, including age and gender, and assessed self-reported olfactory ability on a 7-point Likert scale (1: hard to perceive scents - 7: highly sensitive to scents). 
Next, participants were briefed on the study's purpose and procedures.
They then completed a training session, which involved walking a shorter distance in VR and learning how to operate VR controllers to become familiar with the task.
Following that, participants proceeded to the main study. 

In the main session, participants completed a total of 40 trials (2 device types $\times$ 4 scent alignment conditions $\times$ 5 repeats). The task order was fully randomized to control for order effects.
In each trial, participants began at a designated starting position and \sy{walked forward along a yellow path on the floor (Fig.~\ref{fig_studyspace}-B).} 
They were instructed to stop when they believed they had reached the scent source. 
\sy{Participants were allowed to move freely forward and backward along the yellow path before submitting their response.
}
Once they reached the position where they believed the scent originated in VR, they pressed the `\textit{A}' button on the right controller and returned to the starting point.
Then, a screen appeared in front of them, prompting a question to assess their confidence in their decision, 
\textit{``How confident are you that the position where you stopped is the exact source of the scent?''}
Once participants answered the question, the trial ended. 

\ic{Before the next trial, participants remained stationary for a 30-second ventilation period, during which the wearable device's exhaust fan was activated to reduce residual scent. This interval also served as a short break between trials. After completing 20 trials with the first device type, participants took a 5-minute break before beginning the second device condition. During this break, the classroom door was opened and an external air circulator was operated to further reduce residual scent before the second device condition.}




\subsection{Apparatus}\label{sec_apparatus}

We used a Meta Quest 2, a desktop to run the VR scene, and two types of olfactory devices. Meta Quest 2 has a 104$^\circ$ horizontal and 98$^\circ$ vertical field of view (FoV), with a resolution of 1832 \texttimes{} 1920 pixels per eye and a maximum refresh rate of 120 Hz.  The virtual scene for the study was developed in Unity 3D (version 2022.3.3f1) and 
executed on a Windows 11 desktop with an AMD Ryzen 7 5800H CPU, 16GB of RAM, and an Nvidia GeForce RTX 3070 graphics card. 

Two olfactory delivery configurations were evaluated: \sy{a wearable device and a stationary device.} 
The \sy{wearable} device was used in both Study 1 and Study 2, whereas the stationary device was used only in Study 1. 
Both devices were fabricated via 3D printing and controlled by an Arduino Nano microcontroller~\footnote{The 3D print files for both \sy{wearable and stationary} devices are available in \url{https://anonymous.4open.science/r/OlfactoryDevice-3E5B/}}.
Scent release was achieved by blowing air across a cotton pad infused with the citrus-based orange fragrance oil (2 ml total per session). 
Real-time activation was controlled through serial communication between the Arduino Nano and the Unity application. \ic{The activation latency of approximately 10 ms refers only to the electronic response of the device and should not be interpreted as the time required for airflow or odor to reach the participant.}

The \sy{wearable} device (Fig.~\ref{fig_olfactoryDevice} right) was employed based on prior work~\cite{myung2023enhancing}. 
The device was 5V powered via the headset's USB-C port and weighed approximately 250g. 
It integrates five 5V 50 mm fans: four for scent dispersion and one central exhaust fan used between trials to minimize residual odor accumulation. 
\sy{
Each scent dispersion fan can deliver a scent positioned approximately 5 cm in front of the participants’ nose, providing proximal and consistent olfactory stimulation. 
The fan was electronically activated after participants reached the scent position. 
\sy{Study 1 used only one scent fan so that the wearable device provided a single scent source comparable to the stationary device. This controlled the number of active scent sources across devices, ensuring that differences in localization performance reflected the delivery configuration (wearable vs. stationary) rather than differences in scent presentation.}}

The \sy{stationary} device (Fig.~\ref{fig_olfactoryDevice} left) was adapted from the \sy{wearable} design into a stationary configuration controlled by an Arduino Uno. It was installed at a height of 106.5 cm and angled at 45° toward the participants' face. 
Unlike the proximal delivery of the \sy{wearable} system, the \sy{stationary} device relied on spatial airflow to disperse scent from a distance. 
To compensate for diffusion loss, a larger 12V 92 mm fan was used to project scent across the environment (Fig.~\ref{fig_olfactoryDevice}). 
\ic{
The device was electronically activated when participants reached the predefined scent-source location. }
Between trials, an external air circulator was operated to reduce scent accumulation and cross-condition contamination.

\HDY{
Considering an average ear-to-fan distance of 15 cm for the wearable device with a single operating fan, the wearable device produced a mean sound level of 19.1 dB (range: 16.6–21.3 dB). In contrast, assuming a participant was positioned at the closest perpendicular distance to the stationary device (approximately 50 cm), the stationary device produced a mean sound level of 12.6 dB (range: 8.2–16.8 dB).
To mask operational noise, ambient wind sounds at approximately 32 dB were played in the virtual environment. 
}

\subsection{Measure and Hypothesis}\label{sec_study1_measure}
Four quantitative metrics and one subjective metric were assessed. The quantitative metrics included distance error, error tendency, \ic{Path Efficiency }
 and time per meter, while the subjective metric assessed participants’ confidence level.

\begin{itemize}[leftmargin=0.15in, noitemsep]
    \item {\textbf{Distance Error}}: Evaluates the accuracy of scent localization by calculating the distance between the actual scent source and the participants' indicated location. Values closer to 0 indicate higher spatial accuracy.

    \item {\textbf{Error Tendency}}: 
    Represents the directional bias of distance error. Positive values indicate a tendency to perceive the scent source as farther away than it is, while negative values indicate the opposite. 
     It was calculated by subtracting the distance traveled by participants from the actual scent source position.
\item {\textbf{Path Efficiency}}: Captures participants’ travel efficiency across conditions. Since target locations varied, raw travel distance alone was not directly comparable. Path efficiency was therefore computed
as the ratio of the shortest possible path (Euclidean distance between the start and end points) to the actual path length (distance walked). Values closer to 1 indicate greater travel efficiency.

    \item {\textbf{Time per Meter}}:
    Represents the temporal cost of navigation under each condition. It was calculated as the ratio of completion time to target distance, yielding the average time spent per meter traveled. Lower values indicate faster performance relative to distance.
  
    \item {\textbf{Confidence Rate}}: Measures participants' perceived certainty in their chosen location. After each task, participants rated their confidence on a 7-point Likert scale (1 = not confident at all, 7 = very confident). This measure was used to compare perceived certainty across conditions.
\end{itemize}

Our hypotheses for Study 1 are as follows:

\begin{itemize}[noitemsep]

    \item[\textbf{H1-1:}]The \factor{SAligned} condition is expected to produce the lowest distance error and \ic{highest path efficiency }
     compared to the other conditions (i.e., \factor{SLeading}, \factor{STrailing}, and \factor{SOnly}), as the visual and 
    \sy{scent-delivery feedback} are aligned.
    \item[\textbf{H1-2:}] The \factor{SLeading} condition is expected to produce a positive error tendency (overestimation), as the virtual object appears in front of the actual scent source, while the \factor{STrailing} condition is expected to produce a negative error tendency (underestimation), as the object is placed behind the actual scent source.
    
    \item[\textbf{H1-3:}] The \factor{SOnly} condition is expected to yield higher navigation efficiency than \factor{SLeading} and \factor{STrailing}, but lower efficiency than \factor{SAligned}, as it lacks supporting visual information.
    
    \item[\textbf{H1-4:}]The \factor{\sy{wearable} device} is expected to demonstrate better navigation efficiency (lower distance error, 
    higher path efficiency, and lower time per meter) with a higher confidence rate because it can deliver the scent directly to the participants' nose.

\end{itemize}

\subsection{Participants}

A total of 26 participants were initially recruited through posters and our university's social media platforms. Among them, data from 24 participants (6 males and 18 females; mean age: 22, range: 18–29) were included in the final analysis, excluding 2 participants who withdrew due to cybersickness.
Participants reported no difficulty walking independently, and their self-assessed olfactory sensitivity score had a mean of 5.1 out of 7.
Each participant received a gift card valued at approximately \$7, which met the minimum hourly wage in the country where the study was conducted. 


\subsection{Results}

We analyzed the data using a two-way repeated measures ANOVA 
with a 95\% confidence interval. 
To correct for possible violations of the sphericity assumption, we applied the Greenhouse-Geisser adjustment. 
When significant effects were identified in the ANOVA, we carried out pairwise comparisons using the Bonferroni correction to further examine the differences.
\HDY{Table~\ref{tab:study1_results} presents the summary of significant effects identified in the statistical analysis.
All statistical and descriptive results by Study 1 conditions are reported in Table~\ref{table:study1_statistical_result} and \ref{table:study1_descriptive_result} in the supplemental material.
}
In this section, we focus on reporting the significant effects observed.

\begin{table}[t]
\centering
\small
\caption{Study 1: Significant interaction and main effects.}\label{tab:study1_results}
\begin{tabularx}{0.9\columnwidth}{lXcccc}
\toprule
\textbf{Measure} & \textbf{Effect} & \textbf{df} & \textbf{$F$} & \textbf{$p$} & \textbf{$\eta_p^2$} \\
\midrule
Distance Error & Device $\times$ Scent Location & (3,69) & 2.845 & .044 & .110\\

\midrule
Error Tendency & Scent Location & (3,69) & 32.805 & <.001 & .588\\

\midrule

Path Efficiency & Device Type & (1,23) & 5.116 & .033 & .182\\
& Scent Location & (3,69) & 4.302 & .008 & .158\\

\midrule

Time per Meter & Device Type & (1,23) & 9.551 & .005 & .293\\

\midrule
Confidence & Device Type & (1,23) & 5.471 & .028 & .192\\
\bottomrule
\end{tabularx}
\end{table}

\begin{figure}[t]
  \centering
  \includegraphics[width=.9\linewidth]{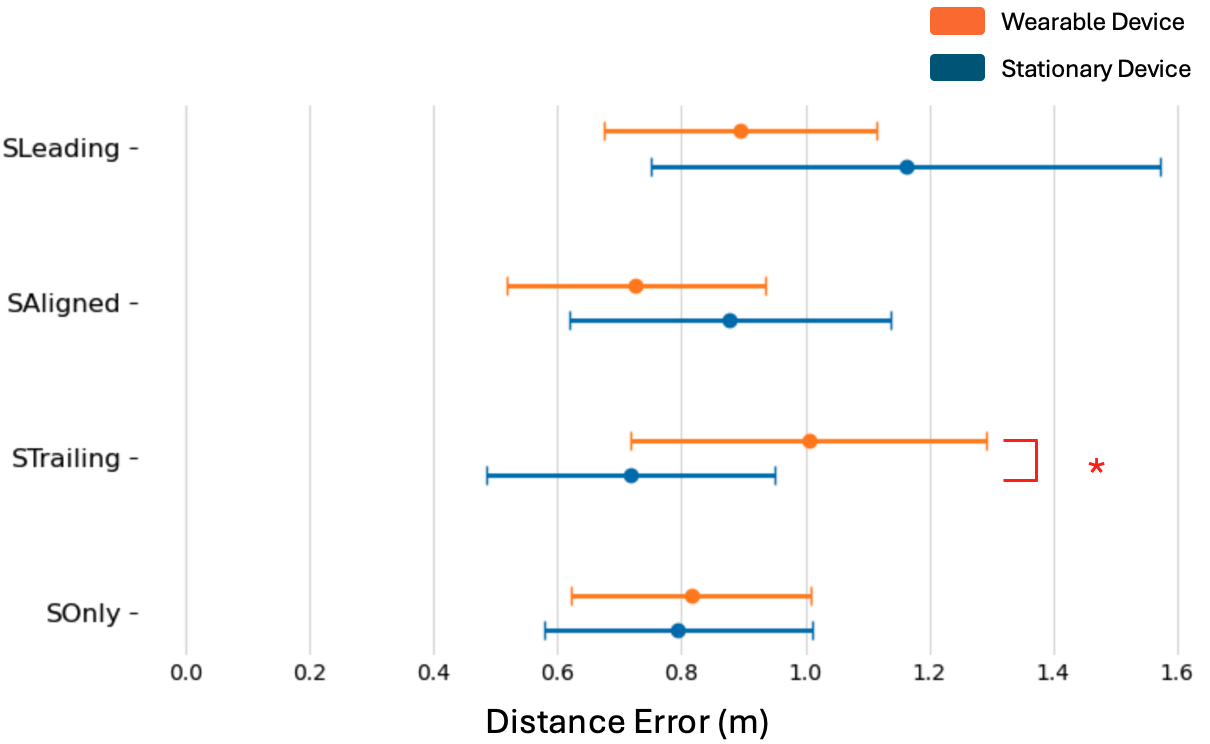}
  \caption{Distance Error across visual-olfactory alignment and device conditions. Error bars indicate 95\% confidence intervals (CIs). 
  }
  \label{fig_study1_distanceError}
\end{figure}



\textbf{Distance Error:} We observed a significant interaction effect between device type and scent location ($p$ = .044). The result is shown in Fig.~\ref{fig_study1_distanceError}.
 A simple effect of device type was found within \factor{STrailing} ($F(1, 23)$ = 4.466, $p$ = .046, $\eta_p^2$ = 0.163).
Participants using \factor{\sy{wearable} device} (M = 1.01, SD = 0.68) had significantly higher distance error than those using \factor{\sy{stationary} device} (M = 0.72, SD = 0.55).


\begin{figure}[t]
  \centering
  \includegraphics[width=.9\linewidth]{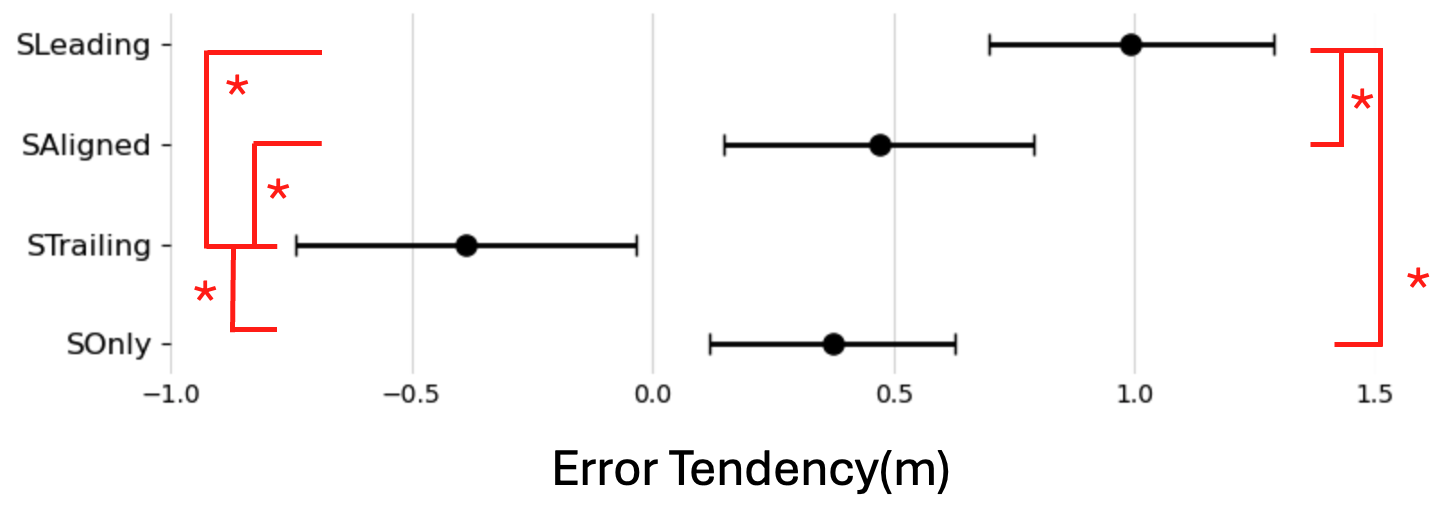}
  \caption{Error Tendency results across visual-olfactory alignment conditions. Error bars indicate 95\% CIs. 
  }
  \label{fig_study1_errorTendency}
\end{figure}

\textbf{Error Tendency:}
The main effect of scent location was found ($p$ \textless .001). 
The results of the scent location conditions were shown in Fig.~\ref{fig_study1_errorTendency}.
Pairwise comparisons showed that \factor{STrailing} (M=-0.39, SD=1.01) had a negative tendency and showed significant differences compared to \factor{SLeading} (M=0.99, SD=0.83, $p$ \textless .001), \factor{SOnly} (M=0.37, SD=0.87, $p$ \textless .001), and \factor{SAligned} (M=0.47, SD=0.86, $p$ \textless .001). \factor{SLeading} had a larger positive error tendency than \factor{SOnly} ($p<.001$) and \factor{SAligned} ($p=.005$). 
\HDY{No differences were observed between wearable (M = 0.35, SD = 0.96) and stationary (M = 0.37, SD = 1.07) device types.}





\begin{figure}[t]
  \centering
  \includegraphics[width=.9\linewidth]{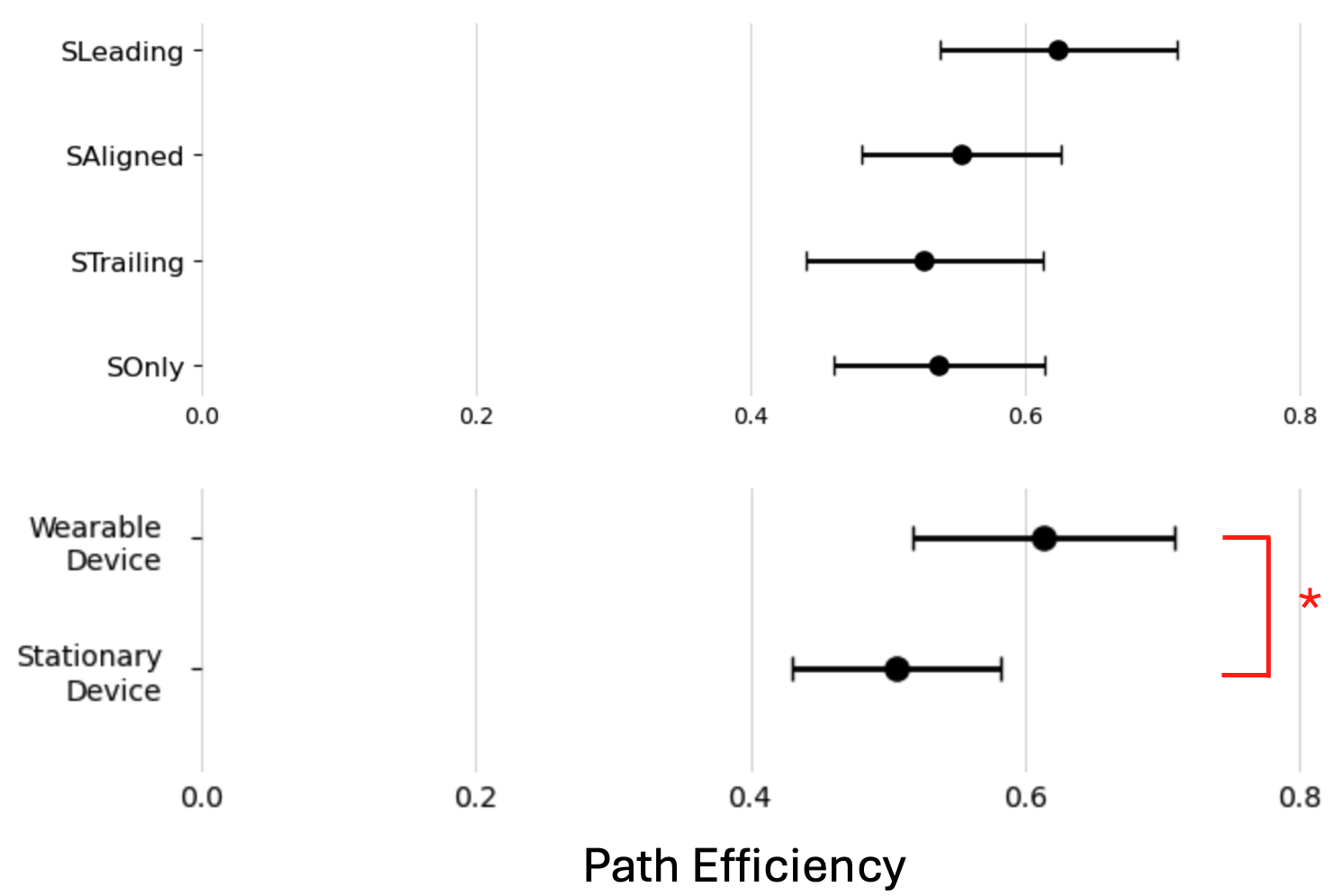}
  \caption{Path Efficiency results by alignment and device conditions, respectively. Error bars indicate 95\% CIs. 
  }
  \label{fig_study1_Tortuosity}
\end{figure}

\begin{figure}[h]
  \centering
  \includegraphics[width=.9\linewidth]{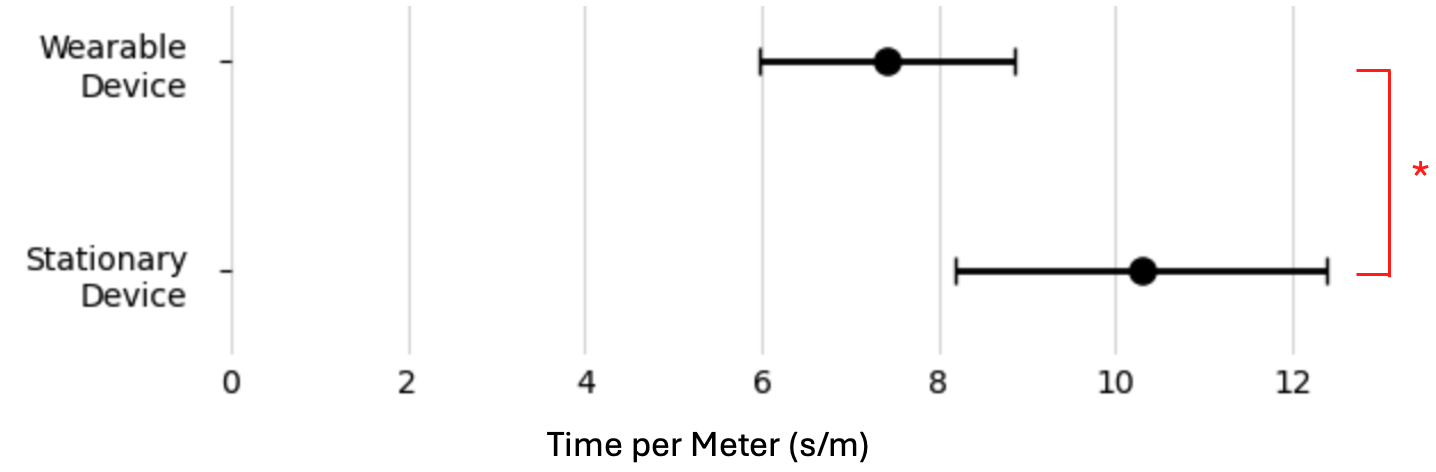}
  \caption{Time per Meter across devices. Error bars indicate 95\% CIs. 
  }
  \label{fig_study1_Timepermeter}
\end{figure}

\textbf{Path Efficiency: }
The main effects of the device type ($p$ = .033) and scent location condition ($p$ = .008) were reported (Fig.~\ref{fig_study1_Tortuosity}).
Participants exhibited 
higher path efficiency when using \factor{\sy{wearable} device} (M = 0.61, SD = 0.26) compared to \factor{\sy{stationary} device} \HDY{(M = 0.51, SD = 0.22)}. 
Between the scent location conditions, however, with the Bonferroni correction, post-hoc comparisons showed no significant differences.



\textbf{Time per Meter:} 
The main effect of device type was found ($p$ = .005). 
Participants reported significantly lower value with the use of \factor{\sy{wearable} device} (M = 7.41, SD = 4.07) than with \factor{\sy{stationary} device} (M = 10.29, SD = 5.4) as shown in Fig.~\ref{fig_study1_Timepermeter}.

\begin{figure}[h]
  \centering
  \includegraphics[width=.8\linewidth]{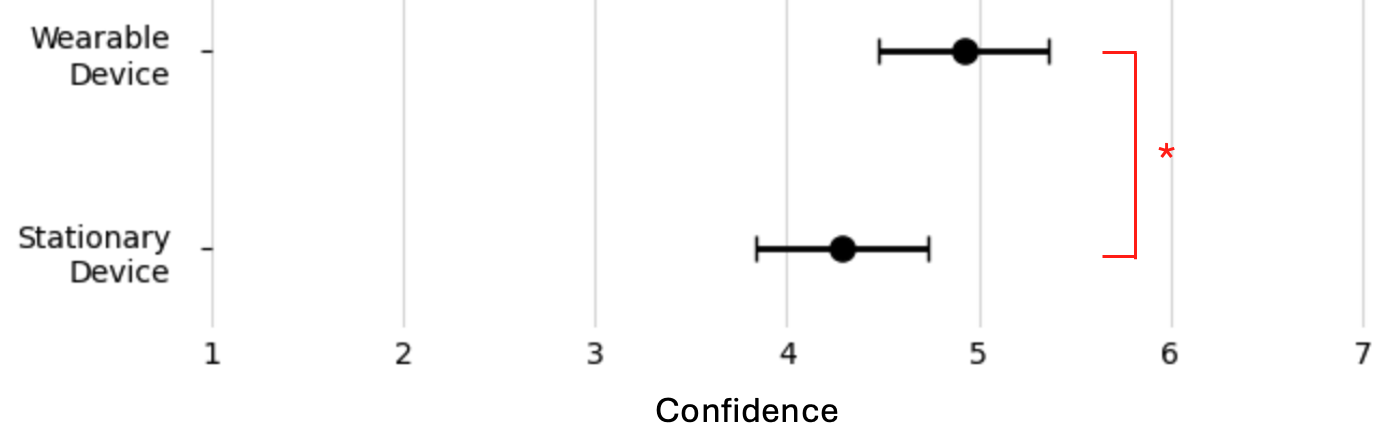}
  \caption{Confidence across devices. Error bars indicate 95\% CIs. 
  }
  \label{fig_study1_Confidence}
\end{figure}


\textbf{Confidence:} 
The main effect of the device type was reported ($p$ = .028), showing significantly higher confidence score for \factor{\sy{wearable} device} (M = 4.92, SD = 1.16) compared to \factor{\sy{stationary} device} (M = 4.29, SD = 1.16). The result is shown in Fig.~\ref{fig_study1_Confidence}.

\subsection{Discussion }
\paragraph{\textbf{Spatial Alignment and Distance Accuracy (H1-1):}} Distance error did not significantly differ across alignment conditions. 
Participants localized the scent source with comparable accuracy whether the visual object was aligned, misaligned, or absent. 
Therefore, \textbf{H1-1} is not supported. 
This result indicates that visual-olfactory congruency did not improve absolute localization accuracy in our walking task. 
One possible explanation is that localization performance was primarily constrained by the spatial discriminability of 
\sy{scent-delivery feedback.} 
Even when visual and 
\sy{scent-delivery feedback} were co-located, the inherent spatial resolution of olfactory perception may have limited further improvements in localization accuracy~\cite{porter2007mechanisms}.

Moreover, participants walked toward the perceived source, which may have helped them update their position during the task ~\cite{zhao2015you}. 
Under these walking-based conditions, coarse olfactory detection together with movement-related cues may have been sufficient for participants to stop at a relatively consistent location.
As a result, spatial congruency did not enhance magnitude accuracy. Its effect appeared more clearly in directional interpretation.

The absence of differences in error magnitude does not imply an absence of cross-modal interaction. 
It indicates that alignment affected how participants interpreted spatial relationships between cues rather than how precisely they could localize the source.

\paragraph{\textbf{Visual Anchoring and Directional Bias (H1-2): }}
Strong support for \textbf{H1-2} was found. 
When visual and 
\sy{scent-delivery feedback} were misaligned, participants exhibited directional bias: \factor{SLeading} produced overestimation, whereas \factor{STrailing} produced underestimation.
Notably, misalignment changed the direction of the error without increasing its overall magnitude. 
This pattern is consistent with reliability-weighted multisensory integration, which predicts that vision often dominates spatial judgments \cite{ernst2002humans, ernst2004merging}. 
When visual and 
\sy{scent-delivery feedback} conflicted, participants appeared to rely more on the visual object than on the scent, showing the visual cue shifted the perceived location of the scent source.



\paragraph{\textbf{Navigation Efficiency and the Role of Visual Support (H1-3):}} 
The expected graded pattern of efficiency, with \factor{SOnly} falling between the congruent and misaligned conditions, was not observed. 
Moreover, navigation efficiency did not differ across alignment conditions. These results do not support \textbf{H1-3}.
The absence of efficiency differences suggests that locomotor behavior in short, linear walking tasks may be relatively robust to cross-modal alignment manipulations. 
Participants approached the scent source with comparable efficiency even when the visual representation of the scent source was absent, indicating that 
\sy{the scent-delivery cue may have been} sufficient to guide forward movement.

As discussed earlier, participants exhibited a directional bias in the misaligned conditions. 
However, this bias did not translate into differences in navigation efficiency. One possible explanation lies in the structured walking path imposed by the straight floor guideline. Because participants followed a fixed trajectory, variability in movement paths was limited, which likely reduced the sensitivity of \ic{path efficiency} 
as a measure of subtle alignment effects. Thus, although misalignment induced directional bias, it did not substantially disrupt locomotor efficiency under these constrained conditions.

\begin{figure*}[t]
  \centering 
  \includegraphics[width=.7\textwidth]{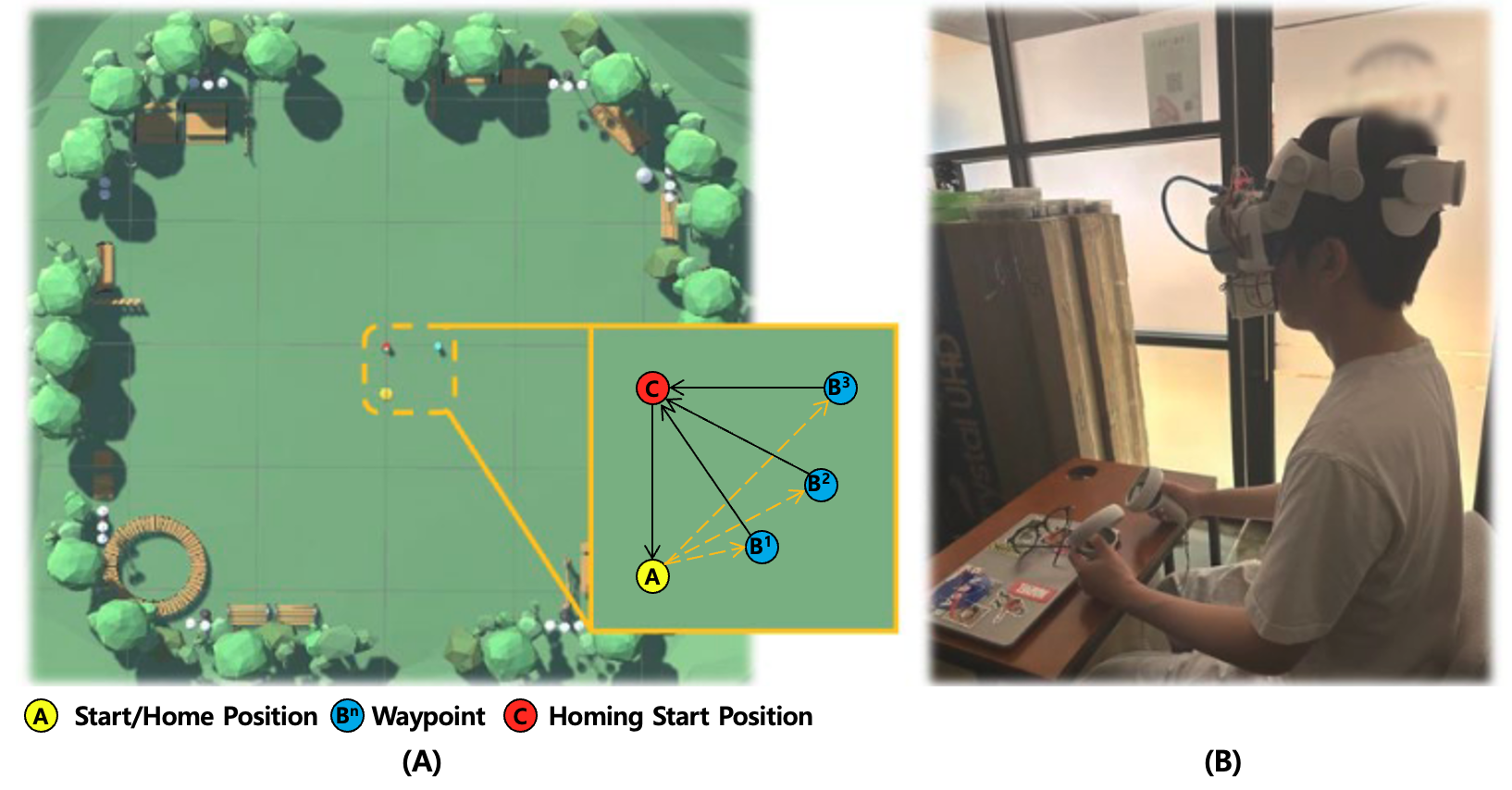}
  \vspace{-.3cm}
  \caption{\sy{(A) VR scene for the TCT in Study 2. 
  Participants started at the yellow pillar, moved to the blue pillar, whose position varied depending on the rotation angle condition, and then proceeded to the red pillar. 
  } 
  Finally, they returned to the starting position \sy{(i.e., the yellow pillar)} to complete the task.
  (B) The \sy{wearable} olfactory device was used in the study. 
\sy{ Ambient wind sounds were played throughout the task to minimize the effect of fan noise.}
Participants performed the task while seated on a chair.
  }\label{fig_study2_task}
\end{figure*}

\paragraph{\textbf{Device Effects and Proximal Scent Delivery (H1-4):}}
Although we found no main effect between the device types for distance error, the \sy{wearable} device yielded greater navigation efficiency, lower time per meter, and higher confidence ratings. 
These results suggest that proximal scent delivery influenced behavioral and decisional responses without improving perceptual accuracy, thereby partially supporting \textbf{H1-4}. 
More specifically, delivering scent closer to the participants’ nose may have produced a more immediately perceived cue and greater certainty about source proximity, thereby reducing hesitation and promoting more decisive movement. 
This distinction between perceptual precision and behavioral efficiency has important implications for olfactory interface design. 

%% file: Section/04_study2.tex
\section{STUDY 2: Homing Task 
}

\subsection{Study Design}

Study 2 employed the TCT task to evaluate how 
\sy{scent-delivery feedback influences homing behavior} and returning accuracy in spatial tasks. 
The original TCT task is designed to evaluate an individual’s spatial navigation and path integration abilities in the real-world~\cite{loomis1993nonvisual}. \ic{Unlike tasks that require users to locate an odor source, Study 2 investigates whether distance-modulated olfactory feedback can support homing toward a remembered starting location.}
TCT has been adapted in VR to study humans' spatial ability and the influence of various sensory modalities on navigation performance \cite{ruddle2011effect, dorado2019homing}. 
As shown in Fig.~\ref{fig_study2_task}, participants are guided along two sides of a triangular path, first from \textit{A} to \textit{$B^n$} and then from \textit{$B^n$} to \textit{C}. 
They are then asked to complete the triangle by returning directly to the starting point without additional guidance.
More details about the task are outlined in Sec.~\ref{sec_study2_task_procedure}.

The study included two independent variables:
\begin{itemize}[leftmargin=0.15in, noitemsep]

\item {\textbf{Scent Presence (\factor{Scent}, \factor{NoScent})}}: 
Participants either received an olfactory cue at the starting location (\factor{Scent}) or no olfactory cue (\factor{NoScent}). In both conditions, participants wore the \sy{wearable} olfactory device (Fig.~\ref{fig_study2_task}B).  We used the same citrus-based orange scent as Study 1 as the target olfactory stimulus.

\item {\textbf{Rotation Angle (\factor{30°}, \factor{60°}, \factor{90°})}}: 
The angle was defined at point C as the angle between segments $B^nC$ and $CA$ (Fig.~\ref{fig_study2_task}A) based on Dorado et al.~\cite{dorado2019homing}.
Although outbound distance varied across the angle condition, the return distance remained constant at 5 m.

\end{itemize}

The virtual environment was designed as a 40m × 40m open space, with minimal visual landmarks such as chairs, tables, and trees placed along the corners of the scene. 
The environment contained only sparse visual landmarks to reduce landmark-based navigation while preserving global orientation cues.~\cite{starrett2021landmarks, sharma2017influence}.
By limiting visual-spatial input, we created a controlled sensory context in which the \sy{effect of distance-modulated scent-delivery feedback} on homing accuracy could be more directly evaluated.




\ic{Participants performed the task while seated and navigated using joystick-based translation and rotation. Specifically, the right joystick controlled translation relative to the HMD heading (maximum 2.0 m/s), whereas the left joystick controlled yaw rotation (maximum 60°/s). Joystick displacement was linearly mapped to movement speed, with motion stopping when the joystick returned to its neutral position.} 
This design choice was motivated by two considerations. First, to reduce the influence of body-motion cues that may have contributed to the results of Study 1, we adopted joystick-based locomotion. Second, this locomotion method reflects a practical constraint in many VR applications, where large-scale physical walking is often not feasible. By minimizing vestibular and proprioceptive input associated with physical locomotion, the design established a more controlled sensory condition for examining olfactory influences on spatial behavior.

\subsection{Study Task and Procedure}\label{sec_study2_task_procedure}

The study lasted approximately 40 minutes. \ic{The study was conducted in a 3.0 m × 4.5 m classroom with no windows. The air conditioning remained turned off throughout the experiment.}
It included the following steps: 1) signing the consent form, 2) completing the pre-questionnaire, 3) receiving a briefing on the study objectives and tasks, 4) performing the training session, and 5) the main session. 


In the main session, participants completed 30 trials (2 scent conditions × 3 triangle angles × 5 repeats). 
Participants performed the TCT task while seated on a chair wearing the \sy{wearable} olfactory device as shown in Fig.~\ref{fig_study2_task}B. The task condition order was randomized to control for order effects.
Within the VR scene, three distinctively colored pillars (yellow, blue, and red) were placed at the center and labeled as A, $B^n$, and C, respectively, as shown in Fig.~\ref{fig_study2_task}A.

\sy{Following the TCT procedure \cite{dorado2019homing}}, participants started at the yellow pillar (Fig.~\ref{fig_study2_task}A-(A)) and moved to the blue pillar (Fig.~\ref{fig_study2_task}A-($B^n$)), \sy{whose location varied according to the rotation-angle condition (i.e., 30°, 60°, and 90°)}. Upon reaching the blue pillar, the yellow pillar disappeared. Participants then proceeded to the red pillar (Fig.~\ref{fig_study2_task}A-(C)). After arriving at the red pillar, they were instructed to return to the original starting position, where the yellow pillar had been located. Regardless of the rotation angle condition, the lengths of the second segment ($B^n$→C) and the return segment (C→A) remained constant at 5 m.
As our primary objective was to evaluate homing accuracy and navigation efficiency, we focused specifically on participants' return behavior. Accordingly, all primary dependent measures were computed during the return segment, defined as the movement from the red pillar (C) to the original starting location marked by the yellow pillar (A).


Depending on the experimental condition, an olfactory cue was presented at the starting location at two different phases: (1) a 3-second burst prior to departure toward the blue pillar, and (2) during the return phase. \sy{Please note that the first scent delivery is only to inform participants that this location corresponded to the home position.}
Olfactory feedback was delivered using the \sy{wearable} device described in Study 1. 
During homing (C $\rightarrow$ A), \sy{the scent-delivery output was modulated according to the participant's distance from the original starting position. 
This modulation was implemented for controlling scent intensity by incrementally increasing the number of active fans in the wearable device from one to four.}
When participants were located within 4m of the starting position, only the first fan was activated. 
As they moved closer to within 3 m, 2 m, and 1 m, two, three, and finally all four fans were activated, respectively.
\sy{This stepwise fan-count modulation was intended to provide distance-dependent scent-delivery feedback rather than a discrete event cue. 
}
\ic{
To minimize the possibility that participants could use fan noise as a cue for localization, 
ambient wind sounds (approximately 32 dB) were continuously played throughout the experiment.
} 
\sy{No scent-delivery fan was activated when the participant was more than 4 m from the starting position.}
\ic{Fan activation was updated continuously based on the participant's current distance from the starting position. Therefore, if participants overshot the target, the number of active fans decreased accordingly as they moved away from the home position.}

When participants believed they had returned to the starting position, they indicated their estimated return location by pressing the A button on the right Meta Quest 2 controller.
After completing a trial, a 30-second ventilation period was used to eliminate residual odor before the subsequent trial.
After completing 15 trials under the first scent presence condition, they took a 5-minute break before proceeding to the next scent presence condition. \ic{During the break, the classroom door was opened, and an external air circulator was used to ventilate the room and reduce residual scent. }


\subsection{Measure and Hypothesis}

We employed three quantitative metrics to evaluate participants’ homing performance. The metrics included distance error, traveled distance, and completion time during the return segment (C $\rightarrow$ A).

\begin{itemize}[leftmargin=0.15in, noitemsep]
    \item {\textbf{Distance Error}}: The Euclidean distance (in meters) between the actual starting position and the location indicated by participants. Smaller values indicate higher homing accuracy.
    
    \item {\textbf{Distance Traveled}}: The cumulative path length (in meters) recorded during the return segment from the red pillar (C) to the participant’s selected stopping location. This metric captures the extent of exploratory movement during homing.

    \item {\textbf{Time Spent}}: The total time (in seconds) was measured from the initiation of movement at the red pillar (C) to the moment when the participant indicated the estimated starting position.
\end{itemize}

Study 2 hypotheses are as follows: 
\begin{itemize}[noitemsep]

    \item[\textbf{H2-1:}]  Distance error is expected to be lower when olfactory cues are present (Scent) compared to NoScent, as olfactory cues can help participants locate the starting position in VR. 
    
    \item[\textbf{H2-2:}] Larger travel angles (90°) will lead to lower distance error than other conditions (30° and 60°) in returning to the original position based on findings of Dorado et al.~\cite{dorado2019homing}.

    \item[\textbf{H2-3:}] Scent will result in shorter distances traveled compared to NoScent, as olfactory cues assist participants in accurately returning to the starting position.

    \item[\textbf{H2-4:}] Scent will lead to faster task completion times than NoScent, for the same reason as stated in H2-3. 
\end{itemize}

\subsection{Participants}

A total of 19 participants (6 male, 13 female, Mean age = 23, range = 18–29 years, SD = 3.65) took part in the study. 
All participants reported no difficulty in perceiving or identifying scents (olfactory sensitivity score M = 4.9, SD = 1.0) during the training session.
Each participant received a gift card valued at approximately \$7. 



\subsection{Results}
\begin{figure*}[t]
  \centering
  \includegraphics[width=.9\linewidth]{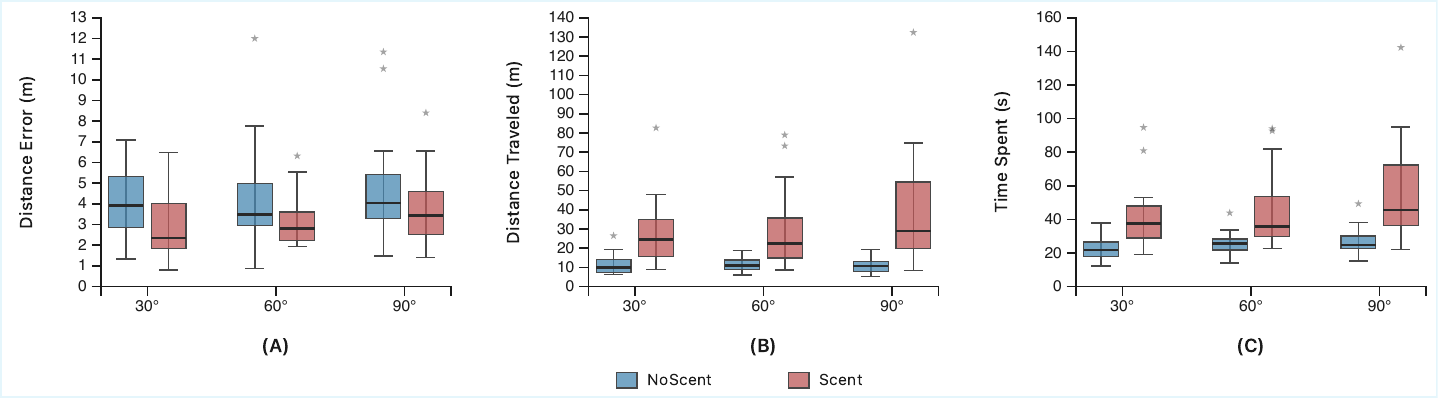}
  \vspace{-0.3cm}
  \caption{Box plots showing the distributions of (A) Distance Error, (B) Distance Traveled, and (C) Time Spent across rotation angles and olfactory conditions. For each rotation angle ($30^\circ$, $60^\circ$, and $90^\circ$), the NoScent and Scent conditions are displayed side by side. Boxes indicate the interquartile range, center lines indicate medians, whiskers extend to 1.5$\times$IQR, and star symbols denote outliers. 
  }
\label{fig_study2_result}
  \vspace{-0.3cm}
\end{figure*}

The same analytical methods as those in Study 1 were employed. \sy{Table~\ref{tab:study2_results} presents the significant effects identified in the statistical analysis, and Fig.~\ref{fig_study2_result} summarizes the descriptive statistics for all study conditions. All statistical and descriptive results by Study 2 conditions are reported in Table~\ref{tab:study2_statistics} and \ref{table:study2_descriptive_result} in the supplemental material.}
This section reports significant effects observed.

\begin{table}[t]
\centering
\small
\setlength{\tabcolsep}{5pt} 
\caption{Study 2: Significant interaction and main effects.}
\label{tab:study2_results}

\begin{tabularx}{\columnwidth}{>{\raggedright\arraybackslash}p{1cm}
                                >{\raggedright\arraybackslash}l
                                cccc}
\toprule
\textbf{Measure} & \textbf{Effect} & \textbf{df} & \textbf{$F$} & \textbf{$p$} & \textbf{$\eta_p^2$} \\
\midrule

Distance Error &
Scent Presence &
(1,18) & 4.486 & .048 & .200 \\

\midrule

Distance &
Scent Presence $\times$ Rotation Angle &
(2,36) & 7.089 & .003 & .283 \\

Traveled &
Scent Presence &
(1,18) & 14.236 & <.001 & .442 \\

&
Rotation Angle &
(2,36) & 7.032 & .004 & .281 \\

\midrule

Time &
Scent Presence $\times$ Rotation Angle &
(2,36) & 8.554 & <.001 & .322 \\

Spent&
Scent Presence &
(1,18) & 20.061 & <.001 & .527 \\

&
Rotation Angle &
(2,36) & 7.031 & .004 & .281 \\

\bottomrule
\end{tabularx}
\end{table}

\textbf{Distance Error:} A significant main effect of \textbf{Scent Presence} was disclosed  ($p$ = .048).
\factor{Scent} (M = 3.3, SD = 1.57) had a smaller distance error than \factor{NoScent} (M = 4.33, SD = 2.35), as shown at Fig.~\ref{fig_study2_result} (A). 

\textbf{Distance Traveled: } Fig.~\ref{fig_study2_result} (B) presents the results for distance traveled. 
A significant interaction effect between \textbf{Scent Presence} and \factor{Rotation Angle} was observed ($p$ = .003). 
A simple effect was revealed that at \factor{30°}($F(1, 18)$ = 12.468, $\eta_p^2$ = 0.409, $p$ = .002). Pairwise comparisons showed that participants traveled significantly longer distances in \factor{Scent} (M = 17.44, SD = 10.34) than in \factor{NoScent} (M = 8.02, SD = 3.98).
We also found a simple effect at \factor{60°}($F(1, 18)$ = 7.126, $\eta_p^2$ = 0.284, $p$ = .016). Pairwise comparisons indicated that distance traveled was significantly greater in \factor{Scent} (M = 22.72, SD = 22.93) than in \factor{NoScent} (M = 8.29, SD = 4.01).
Additionally, a simple effect was found at \factor{90°}($F(1, 18)$ = 17.817, $\eta_p^2$ = 0.497, $p < $  .001). Pairwise comparisons showed that distance traveled was significantly greater in \factor{Scent} (M = 32.12, SD = 23.66) than in \factor{NoScent} (M = 8.35, SD = 3.51).

A significant main effect of \textbf{Scent Presence} was observed ($p$ \textless .001). Participants traveled significantly longer distances in \factor{Scent} (M = 24.092 m, SD = 20.510) than in \factor{NoScent} (M = 8.221 m, SD = 3.773).

A significant main effect of \textbf{Rotation Angle} on traveled distance was found ($p$ = .004). 
Pairwise comparisons indicated that 90° (M = 20.24 m, SD = 20.57) resulted in significantly longer travel distances than 30° (M = 12.73 m, SD = 9.08, $p = 0.012$) and 60° (M = 15.51 m, SD = 17.81, $p = 0.024$). 
\textbf{Time Spent:}  The results for time spent are reported in Fig.~\ref{fig_study2_result} (C). 
A significant interaction effect between \textbf{Scent Presence} and \textbf{Triangle Angle} was found ($p < .001$). A simple effect was found at \factor{30°} ($F(1, 18) = 18.094$, $\eta_p^2 = 0.501$, $p < .001$), with participants in \factor{Scent} (M = 27.27, SD = 17.46) spending significantly more time than those in \factor{NoScent} (M = 11.19, SD = 5.28). A simple effect was also found at \factor{60°} ($F(1, 18) = 14.510$, $\eta_p^2 = 0.446$, $p < .001$), with time spent significantly greater in \factor{Scent} (M = 29.63, SD = 20.8) than in \factor{NoScent} (M = 11.84, SD = 4.11). Additionally, a simple effect was found at \factor{90°} ($F(1, 18) = 20.263$, $\eta_p^2 = 0.530$, $p < .001$), with participants in \factor{Scent} (M = 40.58, SD = 29.39) spending significantly more time than those in \factor{NoScent} (M = 11.21, SD = 3.95).

A significant main effect of \textbf{Scent Presence} was revealed ($p < .001$). 
Participants took significantly longer completion times in \factor{Scent} (M = 32.5, SD = 23.43) than in \factor{NoScent} (M = 11.42, SD = 4.41).


A significant main effect of \textbf{Triangle Angle} was revealed ($p = 0.004$). 
Pairwise comparisons indicated that 90° (M = 25.898, SD = \sy{25.477}) resulted in significantly longer completion times than 30° (M = 19.232, SD = 15.106, $p = 0.018$) and 60° (M = 20.735, SD = 17.322, $p = 0.037$). There \sy{was} no statistical difference between 30° and 60°.

\subsection{Discussion}
\paragraph{\textbf{Effect of Scent on Distance Error (H2-1):}}
Distance error was smaller in the \factor{Scent} condition, indicating that 
\sy{scent-delivery feedback} supported more accurate terminal homing. 
However, this benefit was accompanied by longer traveled distance and completion time.
\sy{The fan-mediated scent-delivery} appears to have provided proximity-related information during the return segment.
\sy{In addition, the stepwise increase in fan activation and scent-delivery output may have provided additional proximity-related feedback} as participants approached the home position. 
This may have enabled participants to refine their positional estimates and reduce terminal localization error.
\sy{However, this finding should be interpreted with caution, as our olfactory device relies on fan-based operation, which may introduce non-olfactory cues. We discuss this limitation further in Sec.~\ref{sec_limitation}.
}



\paragraph{\textbf{Effect of Rotation Angle on Distance Error (H2-2):}}
Distance error did not differ significantly across \factor{30°}, \factor{60°}, and \factor{90°}. 
In contrast to prior findings from full-body TCT paradigms, angular magnitude did not modulate homing accuracy in this joystick-based VR setting. 
One plausible explanation is that joystick locomotion reduced vestibular contributions to path integration, potentially attenuating angular accumulation errors. 
Additionally, the visually sparse virtual environment may have encouraged a consistent navigation strategy across angle conditions. 
Together, these findings suggest that in joystick-based VR contexts with limited embodied input, angular complexity may exert less influence on terminal localization accuracy than in physical locomotion paradigms. 
To further examine this effect, future studies should compare full-body and joystick-based navigation.

\paragraph{\textbf{Effect of Scent on Distance Traveled (H2-3):}}
Participants traveled significantly longer distances in the \factor{Scent} condition, with consistent effects across \factor{30°}, \factor{60°}, and \factor{90°}. 
\sy{Although the distance-modulated scent-delivery} reduced terminal error, it simultaneously increased movement extent during homing.
These results reflect a strategic shift rather than degraded performance. 
In \factor{NoScent}, participants likely relied on ballistic homing, terminating movement once they reached their internally estimated return location. 
When \sy{scent-delivery feedback} was available, \sy{stepwise changes in fan-mediated delivery output} near the target may introduce an opportunity for sensory verification.
\sy{Although we did not systematically record participants' movement trajectories during the task, we qualitatively observed} that participants often slowed down, made fine-grained adjustments, or engaged in correction behaviors to confirm proximity to the home position. 
Thus, the increased travel distance and time spent appear to \sy{reflect confirmation-driven navigation strategy, in which participants made additional exploratory or corrective movements near the target.}

\textbf{Effect of Scent on Completion Time (H2-4):}
Completion time was also significantly longer in \factor{Scent} across all angle levels. The increase in time aligns with the increase in travel distance and further supports the interpretation of a verification-based strategy. 
Rather than accelerating navigation, \sy{scent-delivery feedback} appears to have introduced a more deliberative component near the target location, encouraging fine-grained adjustments before response submission.
\sy{However, we did not systematically record participants’ movement trajectories, limiting our examination of the movement strategies underlying the observed behavior.
We discuss this limitation further in Sec.~\ref{sec_limitation}.
}


%% file: Section/05_overall_discussion.tex
\section{Overall Discussion}
We investigated how olfactory information interacts with visual cues to influence spatial estimation and navigation behavior in immersive VR environments. 
Across two studies, the results demonstrate that 
\sy{scent-delivery feedback} contribute to spatial cognition in VR, but their role differs from that of visual information. 
Olfactory cues did not operate as precise directional signals. Instead, they functioned as supplementary spatial references or proximity-related feedback that influenced how users interpreted spatial relationships and navigated toward target locations. 

First, our findings reveal that visual information continues to dominate spatial interpretation when visual and 
\sy{scent-delivery feedback} conflict. 
In Study 1, spatial misalignment between visual and \sy{scent-delivery feedback} produced systematic directional biases: participants tended to overestimate distances when \sy{scent-delivery feedback} preceded visual objects and underestimate them when \sy{scent-delivery feedback} trailed the objects. 
This pattern suggests that participants anchored their spatial judgments primarily to visual information while integrating \sy{scent-delivery feedback} as secondary signals. 
These findings align with established multisensory integration theories indicating that sensory modalities with higher spatial reliability, such as vision, tend to dominate perceptual judgments when cross-modal conflict occurs.

Second, the results indicate that \sy{scent-delivery feedback} can function as spatial references during navigation.  In Study 2, distance-modulated olfactory feedback reduced terminal homing error in the TCT, suggesting that it provided an additional proximity-related reference when participants approached the goal.
At the same time, the navigation behavior observed in Study 2 reveals a trade-off between spatial precision and efficiency. 
Although distance-modulated olfactory feedback improved terminal homing accuracy, it also led to longer travel distances and increased completion times, suggesting that participants adopted a different navigation strategy when an olfactory cue was available.
Olfactory cue did not accelerate navigation but instead encouraged more deliberate movement aimed at confirming spatial accuracy.

In both studies, we assessed simulator sickness and perceived workload using the Simulator Sickness Questionnaire (SSQ) \cite{kennedy1993simulator} and NASA Task Load Index (TLX) \cite{hart1988development}. They were not reported separately for each study because no significant differences were observed across experimental conditions. These results indicate that the olfactory manipulations and device configurations did not introduce additional discomfort or cognitive workload during the tasks. This finding is important for interpreting the behavioral results, as the observed differences in navigation performance and spatial accuracy are unlikely to be attributable to variations in physical discomfort or task difficulty. Instead, the effects reported in this work more plausibly reflect changes in spatial perception and navigation strategies induced by \sy{scent-delivery feedback} rather than simulator sickness or workload.

Overall, our results suggest that \sy{scent-delivery feedback} plays a complementary role in navigation.  While visual information provides precise spatial structure for orientation and direction estimation, \sy{scent-delivery feedback} can augment this process by offering contextual feedback that helps users verify their position relative to a target location. Importantly, the results demonstrate that \sy{scent-delivery feedback} remains useful even when visual information is absent or partially misaligned, highlighting its potential value in environments with sparse or unreliable visual landmarks.

\subsection{Design Considerations}

Our findings suggest several implications for VR design:

\begin{itemize}[leftmargin=0.15in, noitemsep]

\item \textbf{Use scent as a spatial reference:} \sy{scent-delivery feedback} can serve as effective goal markers, particularly in environments with sparse visual landmarks.

\item \textbf{\sy{Consider wearable} delivery for dynamic interaction:} \sy{Wearable} systems provide flexible and movement-synchronized scent feedback, supporting user confidence and immersion.

\item \textbf{Manage cross-modal alignment carefully:} Visual–olfactory misalignment may not increase accuracy but can induce systematic directional biases.

\item \textbf{Consider the precision–efficiency trade-off:} Distance-modulated olfactory feedback can enhance terminal spatial precision but may increase navigation time and movement extent. It is therefore most appropriate for scenarios where accuracy is prioritized over speed, such as training, rehabilitation, or accessibility support.

\end{itemize}

\subsection{Limitation and Future Work}\label{sec_limitation}
We identified several limitations that also point to directions for future work. 
\sy{First, the fan-mediated scent-delivery system may have introduced non-\sy{olfactory cues}, such as subtle airflow, fan noise, or mechanical awareness, in addition to the intended scent stimulation.
Although we attempted to minimize these effects by presenting ambient wind sounds throughout the experimental tasks, their contribution cannot be fully excluded.
For example, in Study 2, additional fans were activated as participants approached the target location to increase scent intensity. \ic{ We measured the operating sound levels of both devices and continuously presented ambient wind sounds (approximately 32 dB) to mask fan noise. Nevertheless, we cannot completely exclude the possibility that residual auditory or airflow cues contributed to participant performance.} 
While this design enhanced the olfactory signal, it may also have increased airflow or fan noise, allowing participants to use tactile or auditory cues to infer the target position.
}
\sy{Future studies should directly measure device-generated sound levels and further improve auditory isolation while employing more precise scent-delivery methods to better separate the effects of olfactory stimulation from other device-related signals.} 

Second, olfactory intensity was not directly measured during the tasks. 
Because the effectiveness of an olfactory cue may depend on how strongly it is perceived, variation in experienced scent strength across scent types, trials, or device configurations may have influenced localization and navigation behavior.
\sy{Moreover, increasing the number of active fans in Study 2 does not necessarily produce a proportional increase in the intensity of the emitted scent or the intensity perceived by participants.
Future work should therefore incorporate a scent intensity monitoring module into the device to measure and regulate the intensity of the emitted scent more precisely, enabling more consistent control and better understanding of how scent strength influences spatial judgment and movement behavior.
}


Next, the two studies used different locomotion modalities because they targeted different aspects of spatial behavior. 
Study 1 examined localization during physical walking, whereas Study 2 focused on return navigation under controlled joystick-based movement. 
While this design allowed us to investigate scent-delivery feedback across tasks with different spatial demands, it limits direct comparison between the two studies.
Future work should examine similar scent-delivery manipulations under matched locomotion conditions to better isolate the role of movement-related cues and clarify how locomotion shapes the contribution of scent-delivery feedback to spatial behavior.

\sy{Fourth, our analysis of the effects of scent-delivery feedback on navigation behavior in Study 2 is limited.
Although we identified a trade-off between improved homing accuracy and reduced navigation efficiency, the movement strategies participants used to correct their position while searching for the target location remain unclear.
Future work should record continuous movement trajectories to better characterize these movement strategies.
In addition, comparing discrete, constant, and distance-modulated scent-delivery conditions would provide a deeper understanding of how different forms of scent-delivery feedback influence users' navigation behavior.
}
\sy{Future studies should also independently manipulate scent cues presented before navigation and during return navigation to distinguish the contributions of associative spatial memory and online proximity feedback.}

Finally, the participant samples were not fully balanced in gender and age distribution. Because this study was not designed to examine demographic effects, the generalizability of the findings should be interpreted with caution. Future work should test these effects in more demographically balanced samples and examine whether similar patterns are observed across broader participant groups.


%% file: Section/06_conclusion.tex
\section{Conclusion}

This paper investigated how olfactory information interacts with visual cues to influence spatial estimation and navigation behavior in immersive VR environments. Across two user studies, the results indicate that \sy{scent-delivery feedback} contribute to spatial cognition in a complementary role and do not primarily provide precise directional guidance. However, they support navigation by offering contextual feedback that helps users confirm proximity to goal locations. These results highlight the potential of integrating olfactory feedback into immersive systems, particularly in scenarios where spatial verification and environmental awareness are important.

%% file: Section/0_supplemental_material.tex
\clearpage
\onecolumn

  \centering
  {\textsf{\huge Smelling the Way: Olfactory Modulation of Spatial Estimation and
Path Integration in Virtual Reality}}
  \vskip 5pt
  \large \textsf{Supplemental Material}
  \vskip 10pt
\raggedright This document provides additional results beyond the material that we could include in the main paper due to space limitations.

\renewcommand{\figurename}{Fig.}
 \renewcommand\thefigure{\arabic{figure}}  
 \setcounter{figure}{0} 
\renewcommand{\thesection}{\Alph{section}}
\setcounter{section}{0}


\bgroup

\begin{table}[h]
\centering
\caption{Study 1 statistical analysis results.}
\label{table:study1_statistical_result}

\begin{tabular}{llccc}
\toprule
\textbf{Measure} &
\textbf{Factor} &
\textbf{$F$} &
\textbf{$p$} &
\textbf{$\eta_p^2$} \\
\midrule

\multirow{3}{*}{\textbf{Distance Error}}
& Device Type $\times$ Scent Location
& $F(3,69)=2.845$ & .044 & .110 \\

& Device Type
& $F(1,23)=0.176$ & .679 & .008 \\

& Scent Location
& $F(3,69)=1.439$ & .239 & .059 \\
\midrule

\multirow{3}{*}{\textbf{Error Tendency}}
& Device Type $\times$ Scent Location
& $F(3,69)=0.718$ & .545 & .030 \\

& Device Type
& $F(1,23)=0.023$ & .881 & .001 \\

& Scent Location
& $F(3,69)=32.805$ & $<.001$ & .588 \\
\midrule

\multirow{3}{*}{\textbf{Path Efficiency}}
& Device Type $\times$ Scent Location
& $F(3,69)=0.445$ & .721 & .019 \\

& Device Type
& $F(1,23)=5.116$ & .033 & .182 \\

& Scent Location
& $F(3,69)=4.302$ & .008 & .158 \\
\midrule

\multirow{3}{*}{\textbf{Time per Meter}}
& Device Type $\times$ Scent Location
& $F(3,69)=2.768$ & .056 & .107 \\

& Device Type
& $F(1,23)=9.551$ & .005 & .293 \\

& Scent Location
& $F(3,69)=0.387$ & .731 & .017 \\
\midrule

\multirow{3}{*}{\textbf{Confidence}}
& Device Type $\times$ Scent Location
& $F(3,69)=1.811$ & .153 & .073 \\

& Device Type
& $F(1,23)=5.471$ & .028 & .192 \\

& Scent Location
& $F(3,69)=1.111$ & .351 & .046 \\

\bottomrule
\end{tabular}

\vspace{-2ex}
\end{table}

\egroup

\begin{table*}[h]
\centering
\caption{Study 1 descriptive statistics (mean and standard deviation) for each Device $\times$ Condition combination, together with the marginal means for Device and Condition.}
\label{table:study1_descriptive_result}
\begin{tabular}{llccccc}
\toprule
\textbf{Device} & \textbf{Condition} &
\textbf{Distance Error} &
\textbf{Error Tendency} &
\textbf{Path Efficiency} &
\textbf{Time per Meter} &
\textbf{Confidence} \\
& &
\textbf{M (SD) [m]} &
\textbf{M (SD) [m]} &
\textbf{M (SD)} &
\textbf{M (SD) [s/m]} &
\textbf{M (SD) [1--7]} \\
\midrule

\multirow{4}{*}{Stationary}
& SLeading  & 1.16 (0.97) &  1.13 (1.01) & 0.57 (0.28) & 10.94 (5.49) & 4.18 (1.21) \\
& SAligned  & 0.88 (0.61) &  0.46 (0.98) & 0.50 (0.22) & 10.13 (5.61) & 4.46 (1.05) \\
& STrailing & 0.72 (0.55) & -0.43 (0.80) & 0.49 (0.19) &  9.84 (5.37) & 4.20 (1.23) \\
& SOnly     & 0.80 (0.51) &  0.32 (0.90) & 0.46 (0.19) & 10.25 (5.40) & 4.30 (1.18) \\
\cmidrule(lr){1-7}
\multicolumn{2}{l}{\textbf{Stationary Mean}}
& \textbf{0.89 (0.70)}
& \textbf{0.37 (1.07)}
& \textbf{0.51 (0.22)}
& \textbf{10.29 (5.40)}
& \textbf{4.29 (1.16)} \\
\midrule

\multirow{4}{*}{Wearable}
& SLeading  & 0.89 (0.52) &  0.85 (0.59) & 0.67 (0.26) & 6.70 (4.09) & 5.12 (1.06) \\
& SAligned  & 0.73 (0.49) &  0.48 (0.74) & 0.60 (0.21) & 7.25 (4.31) & 4.96 (1.30) \\
& STrailing & 1.01 (0.68) & -0.34 (1.18) & 0.57 (0.29) & 8.55 (4.28) & 4.78 (1.24) \\
& SOnly     & 0.82 (0.46) &  0.42 (0.85) & 0.61 (0.26) & 7.12 (3.57) & 4.83 (1.09) \\
\cmidrule(lr){1-7}
\multicolumn{2}{l}{\textbf{Wearable Mean}}
& \textbf{0.86 (0.55)}
& \textbf{0.35 (0.96)}
& \textbf{0.61 (0.26)}
& \textbf{7.41 (4.07)}
& \textbf{4.92 (1.16)} \\
\midrule

\multicolumn{2}{l}{\textbf{SLeading Mean}}
& \textbf{1.03 (0.78)}
& \textbf{0.99 (0.83)}
& \textbf{0.62 (0.27)}
& \textbf{8.82 (5.25)}
& \textbf{4.65 (1.22)} \\

\multicolumn{2}{l}{\textbf{SAligned Mean}}
& \textbf{0.80 (0.56)}
& \textbf{0.47 (0.86)}
& \textbf{0.55 (0.22)}
& \textbf{8.69 (5.16)}
& \textbf{4.71 (1.20)} \\

\multicolumn{2}{l}{\textbf{STrailing Mean}}
& \textbf{0.86 (0.63)}
& \textbf{-0.39 (1.00)}
& \textbf{0.53 (0.25)}
& \textbf{9.20 (4.85)}
& \textbf{4.49 (1.26)} \\

\multicolumn{2}{l}{\textbf{SOnly Mean}}
& \textbf{0.81 (0.48)}
& \textbf{0.37 (0.87)}
& \textbf{0.54 (0.24)}
& \textbf{8.68 (4.80)}
& \textbf{4.57 (1.16)} \\

\bottomrule
\end{tabular}
\end{table*}

\begin{table}[t]
\centering
\caption{Study 2 statistical analysis results.}
\label{tab:study2_statistics}

\begin{tabular}{@{}llccc@{}}
\toprule
\textbf{Measure} &
\textbf{Factor} &
\textbf{$F$} &
\textbf{$p$} &
\textbf{$\eta_p^2$} \\
\midrule

\multirow{3}{*}{\textbf{Distance Error}}
& Scent Presence $\times$ Rotation Angle
& $F(2,36)=0.203$ & .817 & .011 \\

& Scent Presence
& $F(1,18)=4.486$ & .048 & .200 \\

& Rotation Angle
& $F(2,36)=2.117$ & .135 & .105 \\

\midrule

\multirow{3}{*}{\textbf{Distance Traveled}}
& Scent Presence $\times$ Rotation Angle
& $F(2,36)=7.089$ & .003 & .283 \\

& Scent Presence
& $F(1,18)=14.236$ & $<.001$ & .442 \\

& Rotation Angle
& $F(2,36)=7.032$ & .004 & .281 \\

\midrule

\multirow{3}{*}{\textbf{Time Spent}}
& Scent Presence $\times$ Rotation Angle
& $F(2,36)=8.554$ & $<.001$ & .322 \\

& Scent Presence
& $F(1,18)=20.061$ & $<.001$ & .527 \\

& Rotation Angle
& $F(2,36)=7.031$ & .004 & .281 \\

\bottomrule
\end{tabular}
\end{table}

\begin{table*}[h]
    \centering
    \caption{ Study 2 descriptive statistics (mean and standard deviation) for all combinations}
    \label{table:study2_descriptive_result}
            \begin{tabular}{llccc}
                \toprule
                    \textbf{Scent Presence} & \textbf{Rotation Angle} &
                    \textbf{Distance Error} &
                    \textbf{Distance Traveled} &
                    \textbf{Time Spent} \\
                    & &
                    \textbf{M (SD) [m]} &
                    \textbf{M (SD) [m]} &
                    \textbf{M (SD) [s]} \\
                    \midrule
                    
                    \multirow{3}{*}{NoScent}
                    & 30$^\circ$ & 4.14 (1.70) &  8.02 (3.98) & 11.19 (5.28) \\
                    & 60$^\circ$ & 4.22 (2.76) &  8.29 (4.01) & 11.84 (4.11) \\
                    & 90$^\circ$ & 4.64 (2.56) &  8.35 (3.51) & 11.21 (3.95) \\
                    \cmidrule(lr){1-5}
                    \multicolumn{2}{l}{\textbf{NoScent Mean}}
                    & \textbf{4.33 (2.35)}
                    & \textbf{8.22 (3.77)}
                    & \textbf{11.42 (4.41)} \\
                    \midrule
                    
                    \multirow{3}{*}{Scent}
                    & 30$^\circ$ & 2.90 (1.51) & 17.43 (10.34) & 27.27 (17.46) \\
                    & 60$^\circ$ & 3.27 (1.30) & 22.72 (22.93) & 29.63 (20.80) \\
                    & 90$^\circ$ & 3.72 (1.83) & 32.12 (23.65) & 40.58 (29.39) \\
                    \cmidrule(lr){1-5}
                    \multicolumn{2}{l}{\textbf{Scent Mean}}
                    & \textbf{3.30 (1.57)}
                    & \textbf{24.09 (20.51)}
                    & \textbf{32.49 (23.43)} \\
                    \midrule
                    
                    \multicolumn{2}{l}{\textbf{30$^\circ$ Mean}}
                    & \textbf{3.52 (1.71)}
                    & \textbf{12.73 (9.08)}
                    & \textbf{19.23 (15.11)} \\
                    
                    \multicolumn{2}{l}{\textbf{60$^\circ$ Mean}}
                    & \textbf{3.75 (2.18)}
                    & \textbf{15.51 (17.81)}
                    & \textbf{20.74 (17.32)} \\
                    
                    \multicolumn{2}{l}{\textbf{90$^\circ$ Mean}}
                    & \textbf{4.18 (2.24)}
                    & \textbf{20.23 (20.57)}
                    & \textbf{25.90 (25.48)} \\

                \bottomrule
            \end{tabular}
\end{table*}